\documentclass[aps,prc,preprint,groupedaddress]{revtex4}

\usepackage{graphicx}
\usepackage{amssymb}

\begin{document}


\title{Effects of non-causal artifacts in a hadronic rescattering model for RHIC collisions}


\author{T. J. Humanic}
\email[]{humanic@mps.ohio-state.edu}
\thanks{This work was supported by the U. S. National Science
Foundation under grant PHY-0355007.}
\affiliation{Department of Physics, The Ohio State University,
Columbus, OH 43210}


\date{\today}

\begin{abstract}
It has been shown that calculations based on a hadronic rescattering
model agree reasonably well with experimental results from RHIC
Au+Au collisions. Because of the large particle densities
intrinsically present at the early time steps of Monte Carlo
calculations attempting to model RHIC collisions undesirable
artifacts resulting in non-causality may be present. The effects of
such artifacts on observables calculated from the rescattering model
are studied in the present work in two ways: 1) varying the time
step and 2) using the subdivision method. It is shown that although
non-causal artifacts are present in the rescattering model they have
no appreciable effects on the calculated observables, thus
strengthing the confidence in the results of this rescattering model
for RHIC energies.
\end{abstract}


\maketitle


\section{Introduction}
Calculations from a simple hadronic rescattering model have been
shown to agree reasonably well with experimental results obtained
from the Relativistic Heavy Ion Collider (RHIC) with $\sqrt{s}$ =
130 GeV and 200 GeV Au + Au collisions
\cite{Humanic:2002b,Humanic:2002a,Humanic:2006}. More specifically,
these calculations provide good representations of the data for 1)
the particle mass dependence of the $m_T$ distribution slope
parameters (i.e. radial flow), 2) the $p_T$, particle mass, and
pseudorapidity dependences of the elliptic flow, and 3) the $p_T$,
centrality, and azimuthal angle dependences of two-pion
Hanbury-Brown-Twiss (HBT) measurements. Although the agreement
between the model and data is sometimes more qualitative than
quantitative, this agreement with such a broad range of RHIC
experimental observables is still a noteworthy accomplishment for a
single, simple, model.

Many hadrons are initially produced in a relatively small volume in
RHIC-type collisions, particularly at early times in the
interaction. It is a challenging task for a Monte Carlo calculation
of this type to deal accurately with binary collisions between
particles in such a large particle density environment without
introducing numerical artifacts which may affect the results. One
such undesirable artifact that can occur when the interaction range
between two particles is much greater than the scattering
mean-free-path is non-causality of the collisions resulting in
superluminal particle velocities\cite{Zhang:1998a,Molnar:2000a}. The
method of subdivision can be used to minimize these artifacts
associated with high particle density\cite{Zhang:1998b}.

The effects of such artifacts on observables calculated from the
rescattering model are studied in the present work in two ways: 1)
varying the time step and 2) using the subdivision method. The goal
will be to determine whether non-causal artifacts are present in the
rescattering model and, if so, whether they have appreciable effects
on the calculated observables. This will provide a test of whether
or not one can have confidence in the results of this rescattering
model for RHIC energies. Section II will describe the calculational
methods used and Section III with give the results of the study.

\section{Calculational Methods}
\subsection{Hadronic rescattering calculation}
A brief description of the rescattering model calculational method
is given below. The method used is similar to that used in previous
calculations for lower CERN Super Proton Synchrotron (SPS) energies
\cite{Humanic:1998a}. Rescattering is simulated with a
semi-classical Monte Carlo calculation which assumes strong binary
collisions between hadrons. The Monte Carlo calculation is carried
out in three stages: 1) initialization and hadronization, 2)
rescattering and freeze out, and 3) calculation of experimental
observables. Relativistic kinematics is used throughout.  All
calculations are made to simulate RHIC-energy Au+Au collisions.

The hadronization model employs simple parameterizations to describe the
initial momenta and space-time of the hadrons similar to
that used by Herrmann and Bertsch \cite{Herrmann:1995a}. The initial
momenta are assumed to follow a thermal transverse
(perpendicular to the beam direction)
momentum distribution for all particles,
\begin{equation}
\frac{1}{m_T}\frac{dN}{dm_T}=C\frac{m_T}{\exp{(\frac{m_T}{T})} \pm
1}
\end{equation}
where ${m_T}=\sqrt{{p_T}^2 + {m_0}^2}$ is the transverse mass, $p_T$
is the transverse momentum, $m_0$ is the particle rest mass, $C$ is
a normalization constant, and $T$ is the initial
``temperature parameter''
of the system,
and a gaussian rapidity distribution for mesons,
\begin{equation}
dN/dy=D \exp{[-\frac{(y-y_0)^2}{2{\sigma_y}^2}]}
\end{equation}
where $y=0.5\ln{[(E+p_z)/(E-p_z)]}$ is the rapidity, $E$ is the
particle energy, $p_z$ is the longitudinal (along the beam
direction) momentum, $D$ is a normalization constant, $y_0$ is the
central rapidity value (mid-rapidity), and $\sigma_y$ is the
rapidity width. Two rapidity distributions for baryons have been
tried: 1) flat and then falling off near beam rapidity and 2) peaked
at central rapidity and falling off until beam rapidity. Both baryon
distributions give about the same results. The initial space-time of
the hadrons for $b=0$ fm (i.e. zero impact parameter or central
collisions) is parameterized as having cylindrical symmetry with
respect to the beam axis. The transverse particle density dependence
is assumed to be that of a projected uniform sphere of radius equal
to the projectile radius, $R$ ($R={r_0}A^{1/3}$, where ${r_0}=1.12$
fm and $A$ is the atomic mass number of the projectile). For $b>0$
(non-central collisions) the transverse particle density is that of
overlapping projected spheres whose centers are separated by a
distance $b$. The particle multiplicities for $b>0$ are scaled from
the $b=0$ values by the ratio of the overlap volume to the volume of
the projectile. The longitudinal particle hadronization position
($z_{had}$) and time ($t_{had}$) are determined by the relativistic
equations \cite{Bjorken:1983a},
\begin{eqnarray}
\label{rel} z_{had}=\tau_{had}\sinh{y}\\
 t_{had}=\tau_{had}\cosh{y}\nonumber
\end{eqnarray}
where $y$ is the particle rapidity and $\tau_{had}$ is the
hadronization proper time. Thus, apart from particle multiplicities,
the hadronization model has three free parameters to extract from
experiment: $\sigma_y$, $T$ and $\tau_{had}$. The hadrons included
in the calculation are pions, kaons, nucleons and lambdas ($\pi$, K,
N, and $\Lambda$), and the $\rho$, $\omega$, $\eta$, ${\eta}'$,
$\phi$, $\Delta$, and $K^*$ resonances. For simplicity, the
calculation is isospin averaged (e.g. no distinction is made among a
$\pi^{+}$, $\pi^0$, and $\pi^{-}$). Resonances are present at
hadronization and also can be produced as a result of rescattering.
Initial resonance multiplicity fractions are taken from Herrmann and
Bertsch \cite{Herrmann:1995a}, who extracted results from the HELIOS
experiment \cite{Goerlach:1992a}. The initial resonance fractions
used in the present calculations are: $\eta/\pi=0.05$,
$\rho/\pi=0.1$, $\rho/\omega=3$, $\phi/(\rho+\omega)=0.12$,
${\eta}'/\eta=K^*/\omega=1$ and, for simplicity, $\Delta/N=0$.

The second stage in the calculation is rescattering which finishes
with the freeze out and decay of all particles. Starting from the
initial stage ($t=0$ fm/c), the positions of all particles are
allowed to evolve in time in small time steps (normally $\Delta
t=0.1$ fm/c is used) according to their initial momenta. At each
time step each particle is checked to see a) if it decays, and b) if
it is sufficiently close to another particle to scatter with it.
Isospin-averaged s-wave and p-wave cross sections for meson
scattering are obtained from Prakash et al. \cite{Prakash:1993a}.
The calculation is carried out to 100 fm/c, although most of the
rescattering finishes by about 30 fm/c. The rescattering calculation
is described in more detail elsewhere \cite{Humanic:1998a}.

Calculations are carried out assuming initial parameter values and
particle multiplicities for each type of particle. In the last stage
of the calculation, the freeze-out and decay momenta and space-times
are used to produce observables such as pion, kaon, and nucleon
multiplicities and transverse momentum and rapidity distributions.
The values of the initial parameters of the calculation and
multiplicities are constrained to give observables which agree with
available measured hadronic observables. As a cross-check on this,
the total kinetic energy from the calculation is determined and
compared with the RHIC center of mass energy to see that they are in
reasonable agreement. Particle multiplicities were estimated from
the charged hadron multiplicity measurements of the RHIC PHOBOS
experiment \cite{Back:2000a}. Calculations were carried out using
isospin-summed events containing at freezeout for central collisions
($b=0$ fm) about 5000 pions, 500 kaons, and 650 nucleons
($\Lambda$'s were decayed). The hadronization model parameters used
were $T=300$ MeV, $\sigma_y$=2.4, and $\tau_{had}$=1 fm/c. It is
interesing to note that the same value of $\tau_{had}$ was required
in a previous rescattering calculation to successfully describe
results from SPS Pb+Pb collisions \cite{Humanic:1998a}.

As an indicator of the presence of superluminal artifacts, the
transverse signal propagation velocity, $v_{sT}$, is calculated for
all particle pairs in the rescattering calculation for each time
step from\cite{Molnar:2000a}
\begin{eqnarray}
{\bf v}_{s}= \frac{{\bf r}_1(t_{pres})-{\bf r}_2(t_{prev})}{t_{pres}-t_{prev}}\\
v_{sT}=\sqrt{v_{sx}^2+v_{sy}^2}\nonumber
\end{eqnarray}
where ${\bf v}_s$ is the signal propagation velocity vector, ${\bf
r}_1(t_{pres})$ is the position of particle 1 for the present
collision with particle 2 at time $t_{pres}$, ${\bf r}_2(t_{prev})$
is the position of particle 2 for its previous collision at time
$t_{prev}$, and $v_{sx}$ and $v_{sy}$ are the perpendicular
components of ${\bf v}_{s}$ in the transverse plane. Note that even
though the rescattering calculations are carried out with
relativistic kinematics, it is still possible for $v_{sT}$ to be
greater than $c$ for some particle pairs due to the details of how
the two-particle collisions are implemented.

\subsection{Subdivision method}
The method of subdivision is based on the invariance of Monte Carlo
particle-scattering calculations for a simultaneous decrease of the
scattering cross sections by some factor, $l$, and increase of the
particle density by $l$, where $l$ is called the
subdivision\cite{Zhang:1998a}. As $l$ becomes sufficiently large,
non-causal artifacts become insignificant. The present rescattering
calculation will be tested comparing pion observables from the
``nominal'' calculation, i.e. $l=1$, with subdivisions of $l=5$ and
$l=8$. Note that a subdivision of $l=5$ has been shown to
significantly suppress non-causal artifacts present in some Monte
Carlo calculations\cite{Zhang:1998a,Molnar:2000a}. Since the
particle density increase is accomplished by increasing the particle
number by a factor $l$, the computer CPU time taken per event
increases by a factor $l^2$. For the present study, 1640, 124 and
120 events were generated for the $l=1$, $l=5$, and $l=8$ samples,
respectively. Fortunately, the statistical value of the $l>1$ events
is $l$ times greater per event than for $l=1$. As an example, the
CPU time taken to generate the $l=8$ sample of events was 600
CPU-hours on 2.8 GHz PC processors at the Ohio Supercomputing
Center.

\section{Results}
Figures 1-8 present results from the hadronic rescattering model
simulating hadrons produced in RHIC-energy collisions for various
time steps and subdivisions. Quantities plotted are $v_{sT}$
histograms including all particles in the calculation and several
hadronic observables. Descriptions of how the observables are
extracted from the rescattering calculation are given
elsewhere\cite{Humanic:2002a}. All calculations are carried out for
a mid-rapidity bin, $-2<y<2$, and for an impact parameter of 8 fm to
simulate a medium non-central collision which should result in
significant elliptic flow for the purposes of the present test.
Statistical errors are shown either as error bars or are of the
order of the marker size when error bars are not shown.

\subsection{Time step study}
Figures 1-4 show comparisons of various quantities for three
different time steps in the rescattering calculations, $\Delta
t=0.1, 0.05$ and $0.2$ fm/c. The time step used in calculating
observables which have been compared with RHIC data is $\Delta
t=0.1$ fm/c\cite{Humanic:2002b,Humanic:2002a,Humanic:2006}, so the
other time steps used represent half of the usual step and twice the
usual step. Figure 1 shows $v_{sT}$ histograms.
\begin{figure}
\begin{center}
\includegraphics[width=140mm]{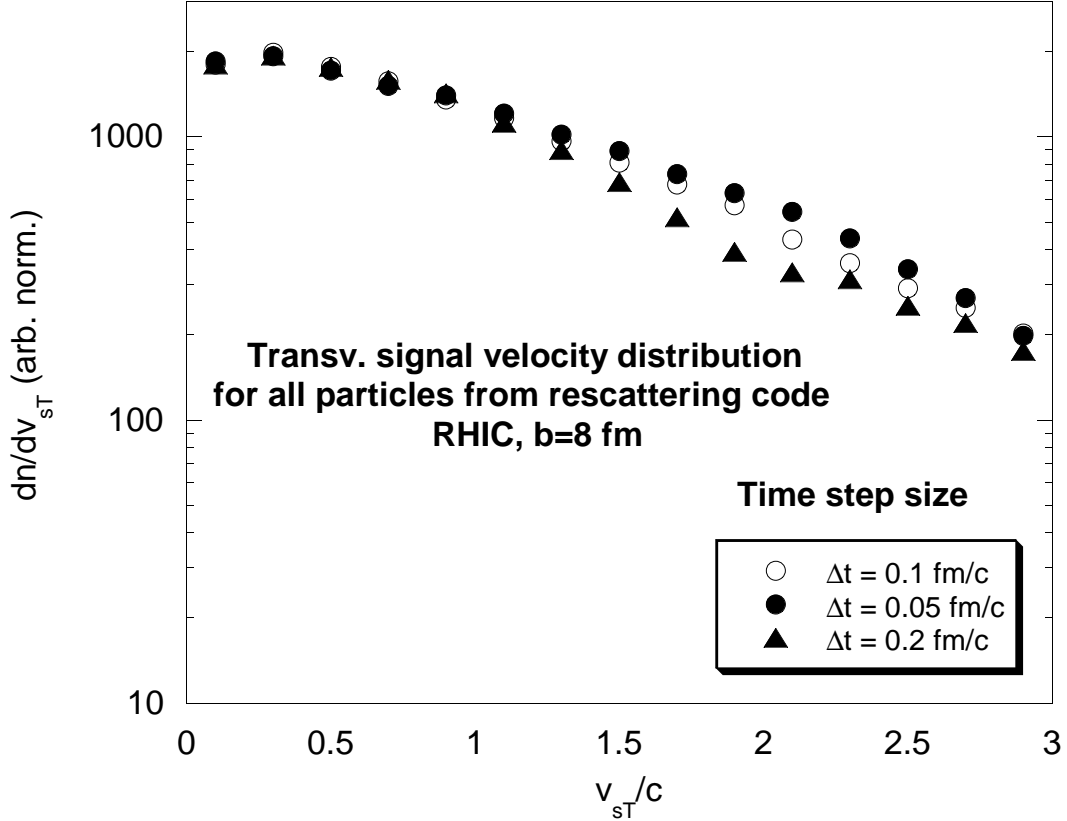}
\caption{\label{fig:sd0} Transverse signal velocity distributions
for all particles for various time steps.}
\end{center}
\end{figure}
It is seen that $v_{sT}$ extends beyond $c$ with an exponential tail
for all time steps, indicating the presence of superluminal
artifacts. For $v_{sT}<c$ all time steps lie on top of each other
whereas for $v_{sT}>c$ they differ somewhat according to what would
be expected for the different time steps, i.e. smaller and thus more
frequent time steps should enhance the superluminal artifacts,
whereas larger and thus less frequent time steps should reduce the
artifacts. For $\Delta t=0.05$ fm/c an enhancement of $\sim 10-20\%$
in the artifacts over the nominal case is seen, while in the case of
$\Delta t=0.2$ fm/c a reduction in the artifacts by about $30\%$
over the nominal time step is seen.

The effects of these time steps on various hadronic observables are
shown in Figures 2-4.
\begin{figure}
\begin{center}
\includegraphics[width=140mm]{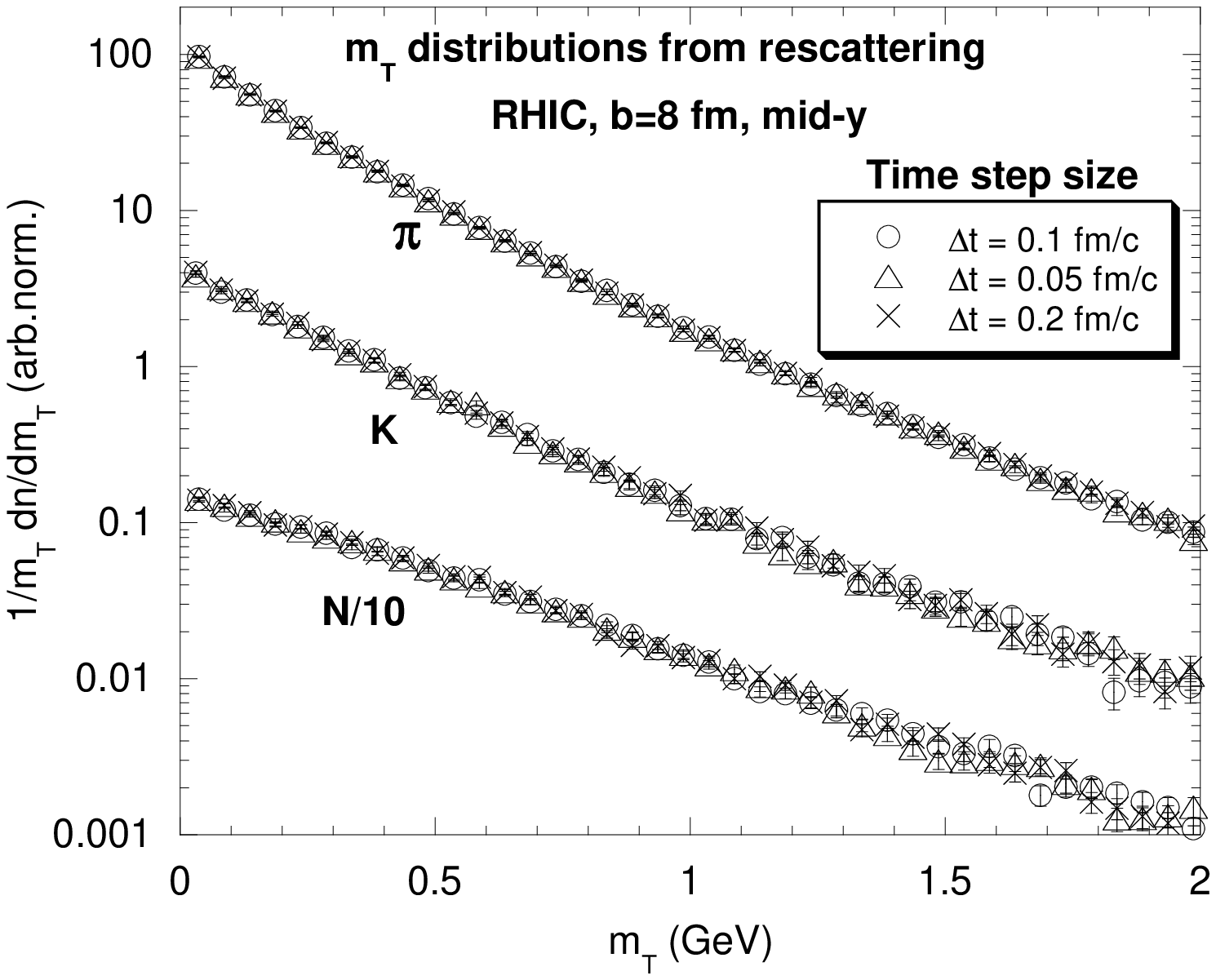}
\caption{\label{fig:sd1} Pion, kaon, and nucleon $m_T$ distributions
for various time steps.}
\end{center}
\end{figure}

\begin{figure}
\begin{center}
\includegraphics[width=140mm]{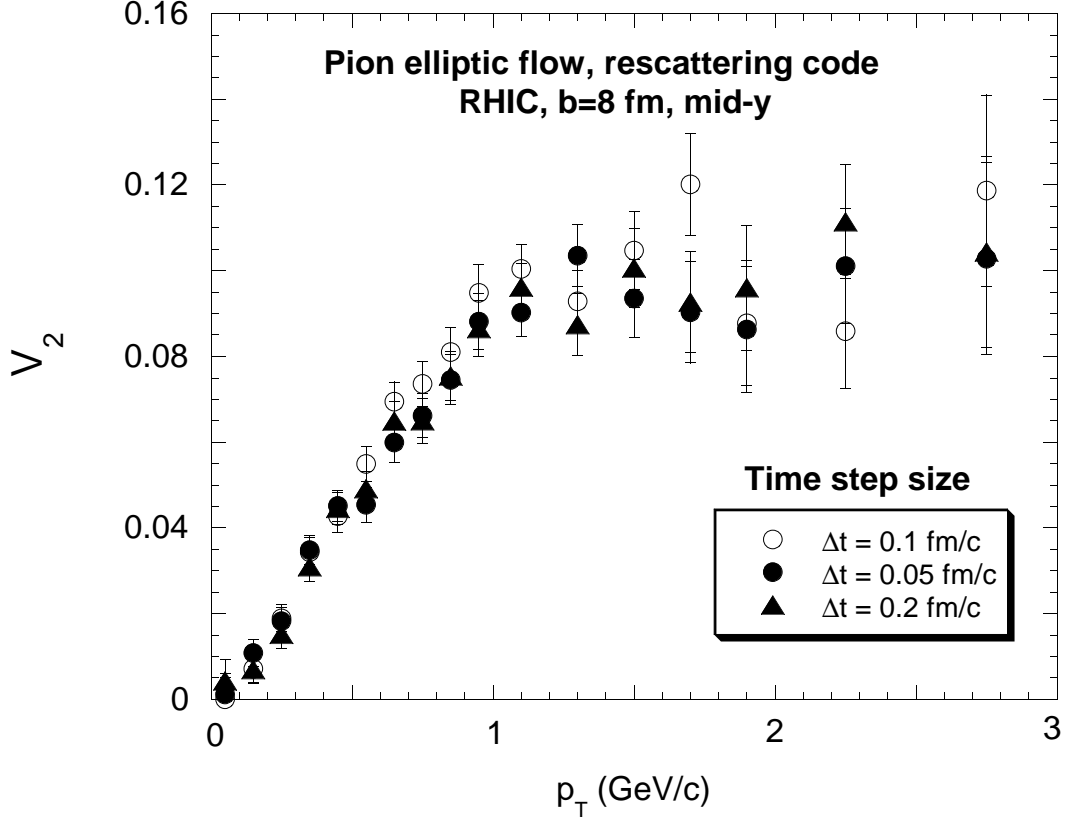}
\caption{\label{fig:sd2} Pion elliptic flow vs. $p_T$ for various
time steps.}
\end{center}
\end{figure}

\begin{figure}
\begin{center}
\includegraphics[width=140mm]{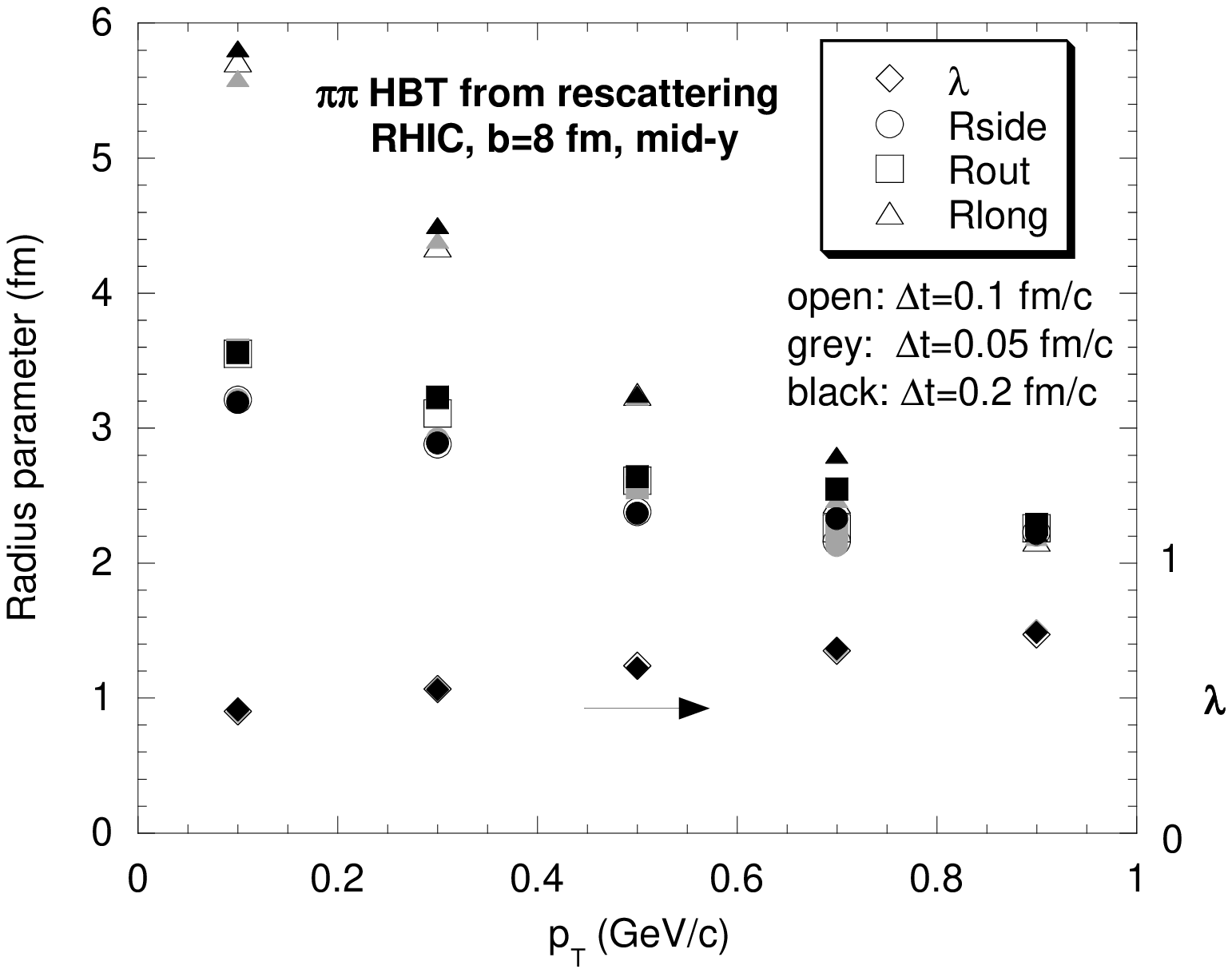}
\caption{\label{fig:sd3} Two-pion HBT source parameters vs. $p_T$
for various time steps. The ordinate scale for $\lambda$ is shown to
the right.}
\end{center}
\end{figure}
Figure 2 shows $m_T$ distributions for pions, kaons, and nucleons,
Figure 3 shows the elliptic flow, $V_2$, vs. $p_T$ for pions and
Figure 4 shows two-pion HBT source parameters\cite{Humanic:2006} vs.
$p_T$. As seen in all three figures, even though the different time
steps have noticeable effects on the artifacts (as seen in Figure
1), they do not significantly effect the hadronic observables, the
observables for all three time steps agreeing within the statistical
errors of the calculations.

\subsection{Subdivision study}
Since it could be argued that the time step study presented above
may not reduce the superluminal artifacts enough to see significant
effects on the hadronic observables, the subdivision test is also
employed since it provides a greater ability to reduce these
artifacts. This is seen in Figure 5 which shows $v_{sT}$ histograms
for the three subdivision used: $l=1, 5$ and $8$.
\begin{figure}
\begin{center}
\includegraphics[width=140mm]{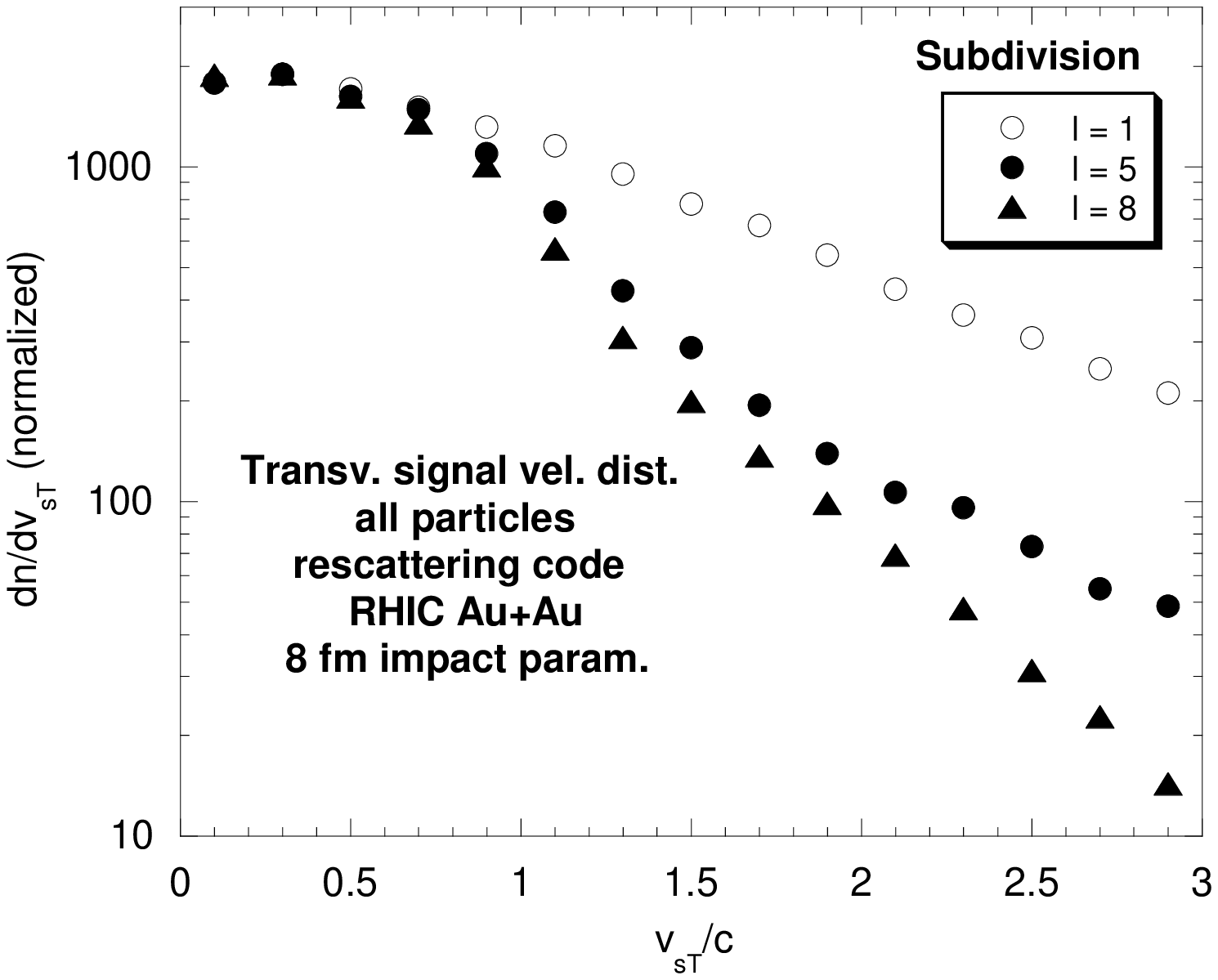}
\caption{\label{fig:sd4} Transverse signal velocity distributions
for all particles for $l=1,5,$ and $8$.}
\end{center}
\end{figure}
While all three subdivisions mostly lie on top of each other for
$v_{sT}<c$, one sees factors of 5-10 reductions in the superluminal
artifacts for $l=5$ and $8$. If these artifacts have a large
influence on the values of the hadronic observables extracted from
the rescattering model, it should surely be evident for the $l=5$
and $8$ cases. Figures 6-8 show the effects of these different
subdivisions on pion $p_T$ distributions, elliptic flow and HBT,
respectively.
\begin{figure}
\begin{center}
\includegraphics[width=140mm]{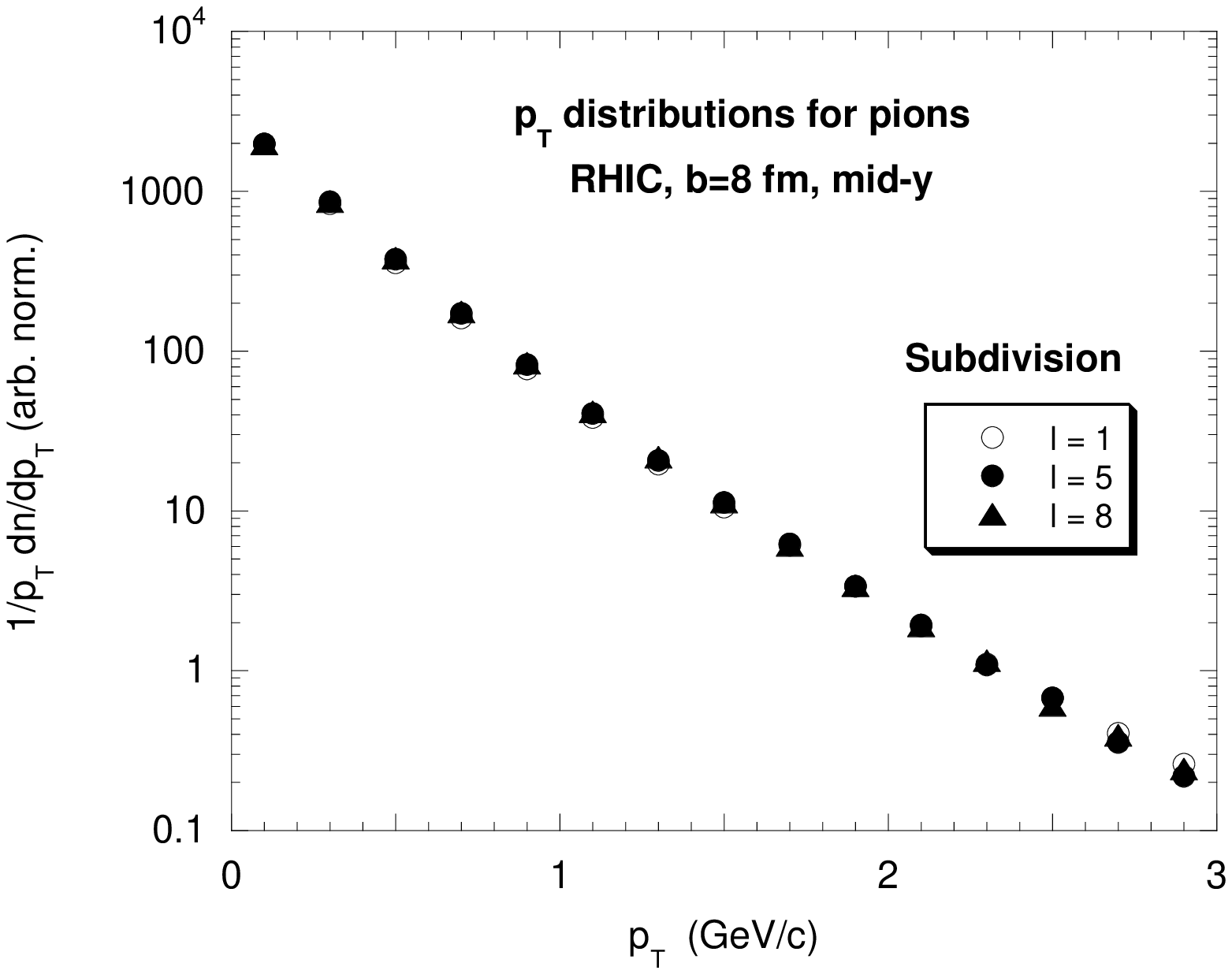}
\caption{\label{fig:sd5} Pion $p_T$ distributions for $l=1,5,$ and
$8$.}
\end{center}
\end{figure}

\begin{figure}
\begin{center}
\includegraphics[width=140mm]{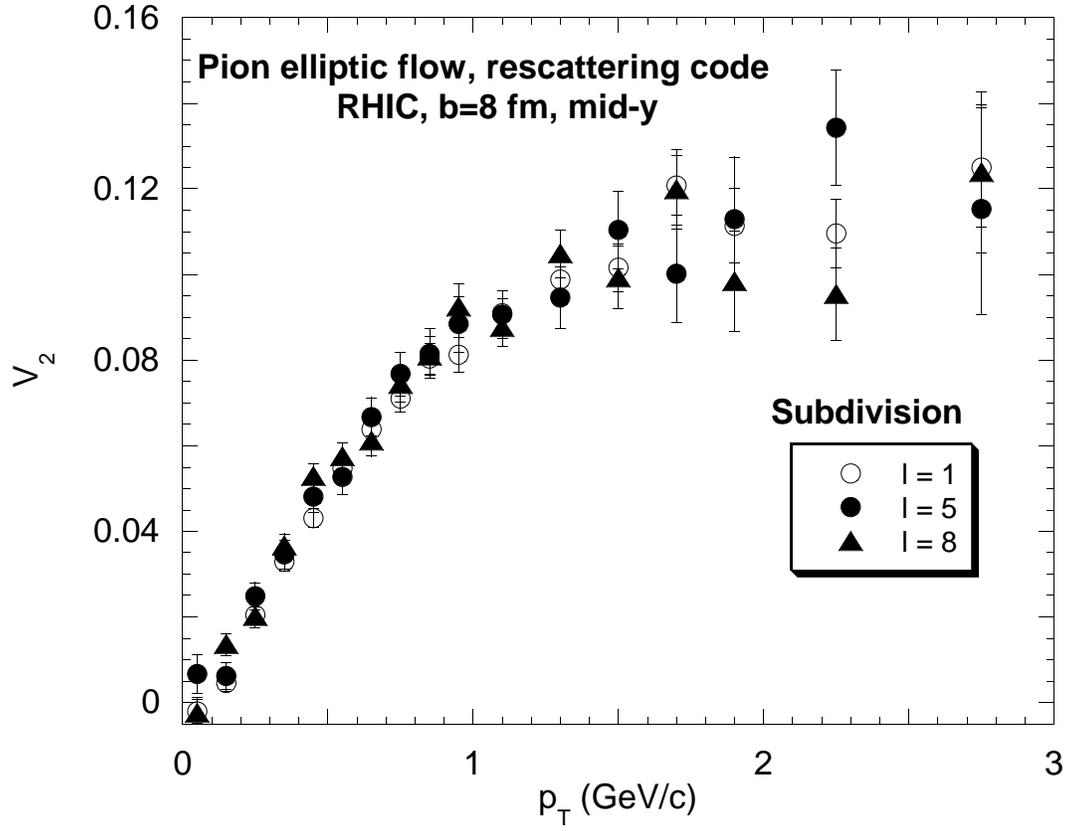}
\caption{\label{fig:sd6} Pion elliptic flow vs. $p_T$ for $l=1,5,$
and $8$.}
\end{center}
\end{figure}

\begin{figure}
\begin{center}
\includegraphics[width=140mm]{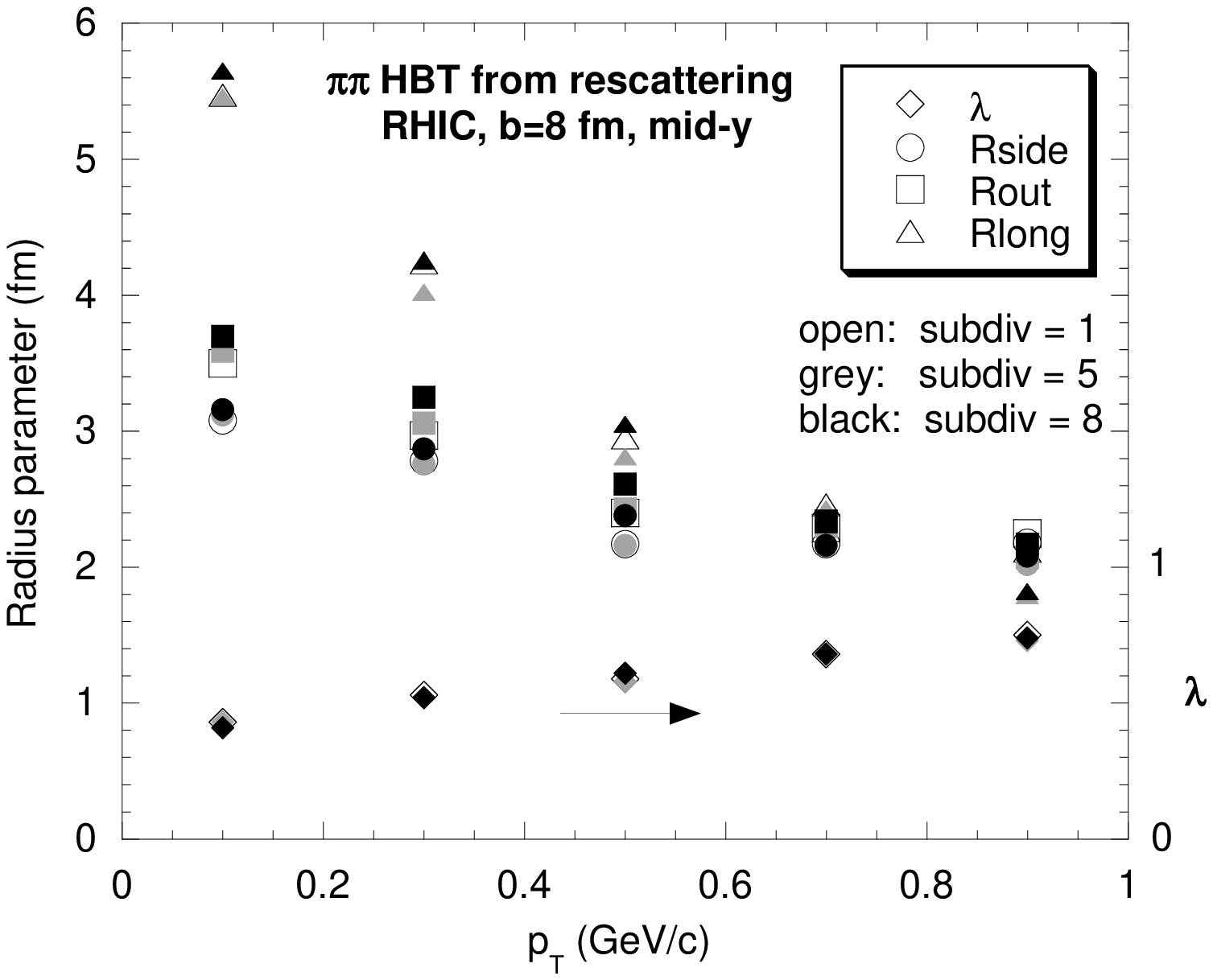}
\caption{\label{fig:sd7} Two-pion HBT source parameters vs. $p_T$
for $l=1,5,$ and $8$. The ordinate scale for $\lambda$ is shown to
the right.}
\end{center}
\end{figure}
As was the case in the time step study, the hadronic observables
shown in Figures 6-8 are not significantly effected by the different
subdivisions used in spite of the fact that superluminal artifacts
are greatly suppressed in the subdivision study.

\section{Discussion and conclusions}
It can be concluded from Figures 1-8 that non-causal (superluminal)
artifacts while being present in the hadronic rescattering code used
in this study do not significantly effect the hadronic observables
calculated from this code. It is possible to speculate on why this
is the case. There are three main features of the code which might
tend to reduce the effects of these artifacts: 1) individual
particles are allowed to scatter only once per time step, 2) a
``scattering time'' of two nominal time steps is defined during
which particles that have scattered are not allowed to rescatter,
and 3) once two particles have scattered with each other, they are
not allowed to scatter with each other again in the calculation. The
present study was performed for non-central collisions ($b=8$ fm),
but it is expected that the same conclusion, i.e. that superluminal
artifacts do not significantly effect the results of calculations
with the rescattering code, would also be obtained for central
collisions. This is because the particle density is not very
different for central collisions in the model than for
mid-peripheral collisions since the particle multiplicities are
scaled by the overlap volume for $b>0$, as mentioned earlier.

In summary, two methods have been used to test the validity of the
present rescattering model calculations in the presence of
non-causal (superluminal) artifacts. It is found that the results of
this model are not appreciably effected by such artifacts, thus
strengthing the confidence in the results presented previously from
this rescattering model for RHIC.

\begin{acknowledgments}
The author wishes to thank Tam\'as Cs\"org\H{o}, D\'enes Moln\'ar
and Derek Teaney for helpful discussions relating to this paper.
This work was supported by the U. S. National Science Foundation
under grant PHY-0355007.
\end{acknowledgments}


\end{document}